\newcommand{\be}{\begin{equation}}
\newcommand{\ee}{\end{equation}}

\documentstyle[epsf,aas2pp4]{article}
\lefthead{Zerbi et al}
\righthead{Search for low instability strip variables in NGC 2516}

\begin{document}

\title{Search for low instability strip variables\\ 
in the young open cluster NGC 2516}
\author{F.M. Zerbi, L. Mantegazza, S. Campana, E. Antonello}
\affil{Osservatorio Astronomico di Brera, Via Bianchi 46, I-23807 Merate, 
Italy.}

\begin{abstract}
In this paper we revise and complete  the 
photometric survey of the instability strip of the southern open cluster NGC 
2516 published by Antonello and Mantegazza (1986). No variable stars with 
amplitudes larger than $0^m.02$ were found. However by means of an accurate 
analysis based on a new statistical method  two 
groups of small amplitude variables have been disentangled: one with periods 
$< 0^d.25$ (probably $\delta$ Scuti stars) and one with periods $>0^d.025$.
The position in the HR diagram 
and the apparent time-scale may suggest that the stars of the second group  belong to a recently discovered new class of variables, named $\gamma$ 
Dor variables. They certainly deserve further study. We also present a  
comparison between the results of the photometric survey and the available pointed ROSAT observations of this cluster.

\end{abstract}

\keywords{Stars: variables: delta scuti -- Stars: variables: other}

\section{Introduction}

The observation of pulsating variables in homogeneous samples such as
stellar clusters is a good way to collect information on the effect of age,
chemical composition and rotation on pulsation.

The studies of open clusters in the northern sky made by several authors (see 
Slovak, 1978) have shown that cluster variable stars of $\delta$ Sct type, 
i.e. the typical low-instability strip pulsators, have an incidence of 
around 30\% apparently independent of age.

The southern open cluster NGC 2516, located at right ascension 
$\alpha(2000)=7^h\: 58^m.3$ and declination $\delta(2000)=-60^o\: 52'$ is 
a young cluster believed to have a common origin with $\alpha$ Per, the 
Pleiades and IC 2606 (Eggen 1983), but unlike these clusters
it has an unusually high abundance of Ap stars.
An extensive study by Mermillod (1981) on young open clusters identifies 
it as the prototype of a group, named the NGC 2516 cluster 
group, that also includes NGC 2168, NGC 2301, NGC 3114, NGC 5460, NGC 6025, 
NGC 7243. They have almost the same age (log age =8.3) and are all 
characterized by a peculiar gap near the turn  off point and by the
presence of numerous Ap stars in the extreme blue region of the main
sequence. 

A first survey for the detection of variable stars in the lower 
part of the instability strip was carried out by Antonello and Mantegazza 
(1986, hereinafter paper I) and provided evidence, besides a normal incidence 
of shorter period variables ($P\leq .25$ d), of a number of longer period 
variables ($P\geq .25$ d), lying on the cool border of the instability strip.

In the last years the presence of variable stars showing small 
amplitudes with longer time scales than the typical $\delta$ Sct 
periodicities has been often reported. Mantegazza et al (1993) first  
gathered a sample of 8 field stars and 5 cluster stars with very similar 
location in the HR diagram, on or just beyond the cool border of the 
instability strip, apparently showing a common behavior. They have been 
named $\gamma$ Dor stars and an updated list has been recently published by 
Krisciunas and Handler (1995).

In the present paper we present previously unpublished observations of 
NGC 2516 that complete the survey for variability inside the instability 
strip of the cluster. The survey is complete in the sense that all the 
stars in the instability strip have been searched for variability.

The absence of a physical lower limit to the possible variability of stars 
implies the need of a statistical method of data analysis in order to
identify the variable stars according to a probability criterion. A method 
was developed and described in paper I.

In this paper we present and use a different statistical approach with 
respect to paper I, less sensitive to the photometric quality of the night.
A comparison between the two approaches is also reported.

\section{Observations} 

A new set of observations were collected at the ESO 1 m reflector from 1987 
January $4$ to January $9$ in the Johnson $B$ filter with the same equipment 
of paper I, i.e. a single channel photometer with a EMI 6256 photomultiplier. 
In the first 4 nights 6 stars per night 
(labeled $s_1-s_6$) were observed following a cycle $s_1-s_2-s_3-s_4-s_5-
s_1-s_2-s_3-s_4-s_6$. The cycle was designed in order 
to observe in each night and for two consecutive nights two of the stars 
already discovered as longer period variables (see 
paper I), namely stars $s_5$ and $s_6$. 
Five stars were observed in each of the last two nights according to the 
cycle $s_1-s_2-s_3-s_4-s_5$ and in this case no previously
known longer period variables were included. 
Each measurement consisted of 25 integrations of 1 second: 
the number of integrations was increased than in paper I in order 
to reduce scintillation noise. Each observed star was monitored for 7.5 
hours and an average number of 85 and 75 points per star have been collected 
in the last two nights and in the first 4 nights respectively.

In order to avoid spurious effects due to variation of air transparency the 
analysis was based on differential time series. Since the difference in 
airmass between the stars observed in each night is very small 
($<0.005$), these spurious effects due to variable air
transparencies are significantly reduced.

\section{Statistical Analysis}

In this run, as for the observations reported in Paper I, no comparison 
stars were assigned {\em a priori}: we then do not know which of the stars 
in each observing cycle are stable or variable. On the other hand the 
expected light variations have such small amplitudes ($<0^m.02$) that
this choice cannot be based on a visual inspection of the differential 
light curves, so a technique had to be developed that extract this 
information from the differential time series between all the possible 
pairs of stars in the cycle.

In paper Ia statistical approach based on
least squares analysis of differential time series was adopted. For the 
$n$ stars in each observing night the variances $s^2_{ij}$ of the 
$n(n-1)$ possible combinations were computed. Under the assumption that 
each of these variances is the sum of the contributions of the two components

\be s^2_{ij}=s^2_i+s^2_j \ee

\noindent the values of such components were estimated by solving the system of 
$n(n-1)$ equations with $n$ unknowns by means of the least squares method.
 
Furthermore the quantities $s^2_i$ were expressed as $s^2_i=a^2_i+b^2_i+c^2$, 
where $a^2_i$ is the {\em intrinsic} variance of the star brightness, 
$b^2_i$ the variance due to the statistical fluctuations of the counts 
(Poisson's variance) and $c^2$ a term which depends on the photometric 
quality of the night. The $b^2_i$ terms could be estimated through the mean 
number of counts collected for each star and subtracted, but the $c^2$ term
remained undetermined and statistics could be performed only on the 
quantities

\be d^2_i=s^2_i-b^2_i=a^2_i+c^2. \ee

Variable stars were discriminated on the basis of an F-test on the quantities $F_i=d^2_i/d^2_{min}$.

However the least squares based procedure may 
not be the best choice. The $n(n-1)$ equations, for instance, are made up of 
pairs of equations, ($i,j$) and ($j,i$) that differ only because in 
the former the times of the $i$-th star have been interpolated to the times 
of the $j$-th star while in the latter vice versa. The number of real 
independent equations is therefore $n(n-1)/2$, resulting in poor
oversampling with respect to the $n$ unknowns. Furthermore if the 
contribution to the variance of the sky and background $c^2$ is appreciable, 
e.g. of the same order of the intrinsic variances, a least squares procedure 
can become unstable.   

We therefore decided to make use of a different approach that 
does not involve the least squares method. Each differential data point in the 
series is defined by 

\be \Delta m_{i,j}(t_k) = -2.5\log_{10}\left({{I_i(t_k)}\over{I_j(t_k)}}
\right) \ee

\noindent where $I_i(t_k)$ and $I_j(t_k)$ are the counts obtained at the 
sampling instant $t_k$, and are therefore 
affected by a statistical error $\sqrt{I_i(t_k)}$ and $\sqrt{I_j(t_k)}$ respectively, due to Poisson's statistics.
Error propagation states that contribution of the statistical error to the 
variance on the single differential measurement is the following

\be s^p_{i,j}(t_k)= 1.179 \left( {{1}\over{I_i(t_k)}}+{{1}\over{I_j(t_k)}} 
\right), \ee

\noindent that can be directly subtracted from measured variances

\be s^2_{i,j}= {{1}\over{N_{i,j}-1}}\left(\sum^{N_{i,j}}_{k=1}
{(\Delta m_{i,j}(t_k)-\bar{\Delta m_{i,j}})^2}\right) -s^p_{i,j}(t_k), \ee

\noindent where $N_{i,j}$ stands for the number of measurements in the given 
time-series.

The Poisson corrected variance may be expressed as 

\be s^2_{i,j}= s^2_i+s^2_j+s^2_r
\ee

\noindent i.e the sum of the components due to each star fluctuation plus an 
additional term due to sky and instrumentation. Indeed the term due to
sky and instrumentation affects directly the differential measurement
and in a first approximation does not depend on the stars observed.
Being corrected for statistical noise contribution and sky background the 
quantities $s^2_i$ may be considered equivalent to the quantities $a^2_i$
better than those named $s^2_i$ in paper I: we will therefore refer 
hereinafter to $s^2_i$ as the {\em intrinsic} variances.
 
The information we need to disentangle variables from constant stars, i.e. 
$s^2_i$, is then distributed into $n-1$ quantities in the form (6) and 
is contaminated by a systematic background error $s^2_r$.

To extract this information we compute the summation over each possible 
differential combination of the $i$-th star

\be S^2_i=\sum_{j\ne i}{s^2_{i,j}}= (n-2) s^2_{i} +\left( \sum^n_{j=1}
{s^2_j} + (n-1) s^2_r \right), 
\ee

\noindent which is a quantity related to the intrinsic variance $s^2_i$. The 
term between parenthesis is constant for all the stars observed in the night.
In order to estimate such a constant term we summed the quantities 
(7) over all the stars observed in the night 

\be S^2=\sum^n_{i=1} S^2_i=2(n-1)\sum^n_{j=1}{s^2_j}+n(n-1)s^2_r.
\ee

Through eqn (7) and (8) we then obtain

\be
s^2_i={{S^2_i}\over{n-2}} -{{S^2}\over{2(n-1)(n-2)}} -{1\over2}s^2_r
\ee

\noindent where the only unknown on the right hand side is the background term 
$s^2_r$.

The observational procedure does not allow a precise estimate of the
background term $s^2_r$. However the minimum value among the differential 
variances obtained in the considered night put an upper limit to this 
quantity, i.e.

\be
s^2_r< min(s^2_{i,j})\;\; i=1,n \;\; j=1,n.
\ee

\noindent Indeed from equation (6) one can see that $s^2_r$ equals the 
minimum value if $s^2_i=s^2_j=0$. The minimum differential variances obtained
range between 1 and 1.5 $10^{-5}$ mag$^2$.     

By introducing this upper limit in  (8) instead of the real unknown
value of background uncertainty, we tend systematically to underestimate 
the intrinsic variances,  the only risk being to assume as constant a 
variable star. For the purposes of this paper such a conservative 
character is welcome.

We then obtained a scale of the background-corrected estimates of the
intrinsic variance for each of the stars observed in a given night.
As in Paper I we could delineate a limit between constant
and variables stars through a statistical F-test by comparing
the intrinsic variances with the minimum value for the night.
 However it should be noticed that the application of an F-test 
here as well as in the context of the analysis of Paper I is not rigorous:
indeed the variances involved are not directly obtained from the
original data, but derived through some manipulations.

For this reason we preferred to assume a fiducial limit of two times 
the minimum intrinsic variance for the night, i.e.  greater 
than the typical F-test value of Paper I. 

We report in tab.~1 the results of the analysis, while a reanalysis by means of the new procedure of  data published in Paper I is reported in tab. ~2.

\newpage
\begin{table}
\begin{center}
\footnotesize
\begin{tabular}{l r c | l r c}
\hline
star&$s_i$ &class     & star&$s_i$&class\\
\hline
C56 &    .529  &C & C40 &    .526  &C            \\
C44 &    .585  &C & C107&    .876    &C          \\
C21 &    .797  &C & C46 &    .885    &C          \\
C42 &    .906   &C & C52 &    1.093    &V(L)     \\
C52 &    1.065    &V(L) & C93 &    4.563    &V(L)\\
C93 &    4.506    &V(S) & C84 &    5.975   &V(S) \\
\hline
C96 &    .309   &C & C98 &    .228    &C         \\
C68 &    .368   &C & C97 &    .328    &C         \\
C28 &    .382   &C & C96 &    .363    &C         \\
C54 &    .623   &V(S) & C55 &    1.163    &V(S)  \\
C69 &    1.822   &V(L) & C106&    2.263    &V(L) \\
C38 &    2.284   &V(S) & C69 &    6.469    &V(L) \\
\hline 
C108&    .290  &C & C66 &    .495   &C           \\
C77 &    .336  &C & C60 &    .535   &C           \\
C59 &    .396  &C & C17 &    .676   &C           \\
C101&    .452  &C & C51 &    1.827    &V(S)      \\
C92 &    1.557 &V(L)&C64&    5.983    &V(L)  \\
\hline
\end{tabular}
\end{center}
\caption{Results of the statistical analysis with the new approach
of the last season data. 
The variances are expressed in $\times 10^{-5}$ $mag^2$. V stays for
variable star while C for constant star according to the classification
reported in the text. L and S means {\em long} and {\em short} period
variables respectively.}
\end{table}

\begin{table}
\begin{center}
\footnotesize
\begin{tabular}{l r c | l r c}
\hline
star&$s_i$ &class     & star&$s_i$&class\\
\hline
C38 &    .320  &C & C46 &    .213    &C          \\
C97 &    .543  &C & C44 &    .237    &C          \\
C54 &    .884  &V(S) & C42 &    .265    &C       \\
C55 &    1.430  &V(S) & C84 &    4.688    &V(S)  \\
C96 &    1.584  &V(L) & C93 &    5.843    &V(L)  \\
\hline 
C109&    .282  &C & C67 &    .705    &C \\
C47 &    .404  &C & C22 &    .104    &C\\
C17 &    .555  &C & C21 &    .113    &C$^*$\\
C60 &    1.103  &V(S) & C51 &    1.736    &V\\
C59 &    1.181  &V(S)& C62 &    3.822    &V(L)$^*$\\
\hline
C4  &    .284  &C & C77 &    .202  &C\\
C99 &    .292  &C & C45 &    .209  &C\\
C32 &    .417  &C & C50 &    .227  &C\\
C31 &    .536  &C & C53 &    .384  &C\\
C69 &    1.699  &V(L) & C52 &    .704  &V(L)\\
\hline
\end{tabular}
\end{center}
\caption{Results of the statistical analysis with the new approach
of the data presented in paper I. The variances are expressed in $\times 10^{-5}$ $mag^2$. V stays for
variable star while C for constant star according to the classification
reported in the text. L and S means {\em long} and {\em short} period
variables respectively. Objects of the first run that
yielded different results with respect to the statistical approach of 
paper I are marked with a star}
\end{table}

A precise estimate of the time-scale of variability for the recognized
variables can not be obtained with such a short baseline. Indeed the 
average baseline of $0.3$ d only allows a poor spectral resolution of 
$3$ d$^{-1}$. However for the purposes of this paper we are only interested 
in distinguishing $shorter$ time-scale variables ($P<.25$ d) and $longer$ 
time-scale variables ($P>.25$ d): this information can in fact be 
extracted even from poor spectral analysis.

We performed a least-squares frequency analysis by means of Vanicek's 
(1971) technique and labeled L the stars that provided evidence of reduction 
factor concentrated at low frequencies ($f<4$ d$^{-1}$) and S all 
the other variables (see columns 3 and 6 of tab. 1 and 2).

\section{Results}

The new set of observations confirmed the variability of 7 stars already 
recognized as variables in the first run, 3 of them labeled L 
(C52, C69 and C93) and 4 of them labeled S (C84, C54, C55, C51). 
A further group of 3 stars (C106, C92 and C64) were pointed out to be 
variables of type L.

As an example of the quality if the light curves obtained during the survey 
differential curves of the shorter period variable C55 and the longer period
variable C69 are reported in fig.~ 1 together with the constant star C97. 
Each curve is computed with respect to C98, the star showing the minimum 
intrinsic variance according to the statistics used.

\begin{figure}
\epsfxsize8truecm
\epsffile{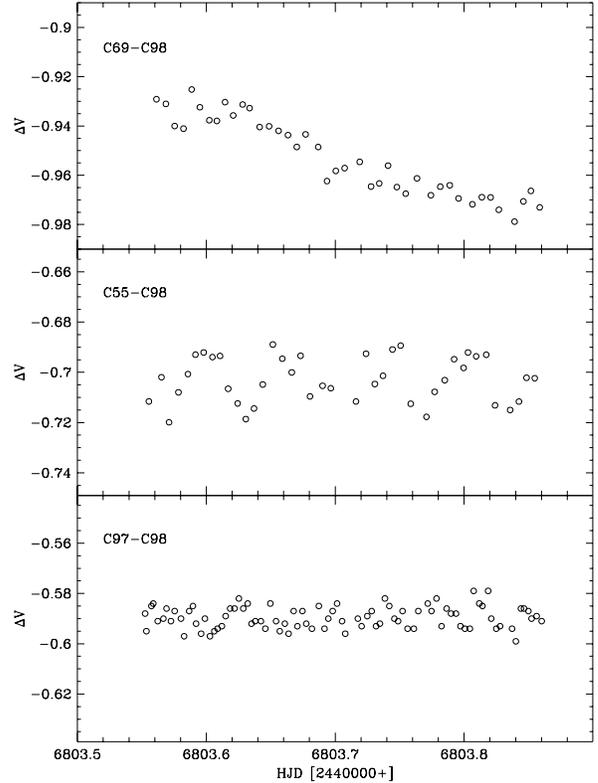}
\caption{Differential light curves of the L-type variable C69 and the 
S-type variable C55 observed in the night of HJD 2446803. The star C98, the
one with the minimum intrinsic variance in the night, was used ascomparison star. The differential light curve of the constant star C97 is also riported for comparison.}
\end{figure}

Only a few discrepancies have been found between the results obtained 
with the old and new approach in the analysis of the first run,
and between the results of the two runs. 
The new analysis of the night HJD 2446115 provided the following different 
results with respect to paper I: the star C21, indicated as an S 
variable, was judged constant by the new statistics, while the star C62, 
previously considered constant, turned out to be an L variable.
Such different results are probably due to the non perfect atmospheric
conditions on that particular night to which the new statistics
is far less sensitive.

The new observations of the stars C96 and C59 do not confirm the 
variability observed in the first run, while C38, constant in the first run,
was classified as S variable in the second run. 
We notice that all the three stars provided evidence of the highest variance
among those observed on the night in which they were classified as variable,
so they were very likely varying when observed for the first time.

In the case of C96 this discrepancy could possibly be due to the fact 
that a longer period variable observed on 
a short baseline near a maximum or a minimum could easily be 
misinterpreted as constant. In the case of the short period variables 
C59 and C38, one possibility is that we are dealing with
multi-periodic pulsators, as most $\delta$ Scuti stars are,
and the apparent constancy is due to a disruptive
beat between different modes.
The possible variability of C38 was already discussed by
Snowden (1975) and Maitzen and Hensberge (1981).

The four newly discovered L type variables (C62, C106, C92 and C64), added
to the group already pointed out in paper I (C93, C52, C96 and C69), make up a 
sample of 8 early F-type, longer time-scale variables.

\section{Discussion}

A sample of 44 A- and early F-type stars has been observed in both runs. This 
sample is complete in the sense that all dwarfs and subgiants with 
$0.00<(b-y)_o<0.25$ present in the cluster have been observed. Among 
these stars, six  were pointed out to be S variables (13\% of the sample, 
excluding the doubtful cases C59 and C38) and 8 to be L variables 
(18\% of the sample). While the incidence of shorter period variables in 
this cluster, as already stated in paper I, can be considered normal,  the
incidence of longer period variables is certainly surprising.

All the stars observed are reported in the HR diagram shown in fig.~2 . The 
stars are plotted with their absolute visual magnitude $V_o$ versus their 
dereddened $(b-y)_o$ color index. The Str\"omgren photometry used as well as 
the evaluation of the color excess $E(b-y)$ and visual magnitude $M_v$ are 
those published by Snowden (1975): main sequence and borders of the 
instability strip are taken from Rodr\'{\i}guez et al (1994). 

\begin{figure}
\epsfxsize8truecm
\epsffile{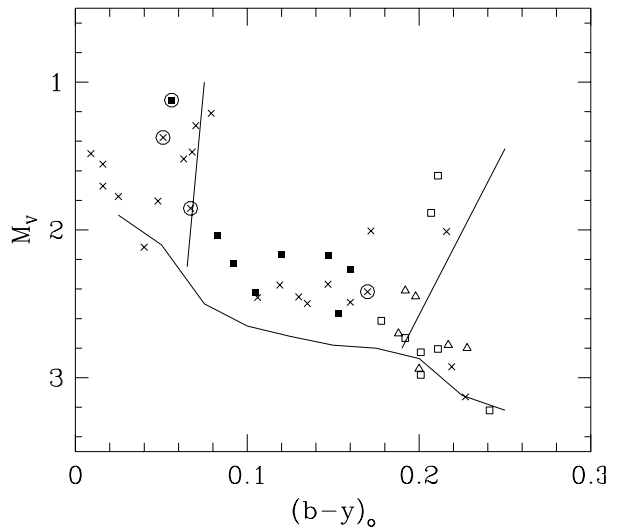}
\caption{Position in the HR diagram for all the stars observed in NGC 2516.
Squares, filled squares and crosses represent L-type, S-type variables 
and non--variable stars respectively: those with the symbol included in an open
circle have been found to be X-ray sources. Field $\gamma$ Dor objects are
represented by triangles}
\end{figure}

With the exception of the peculiar case C 38 all disentangled shorter period 
variables fall within the limits of the $\delta$ Scuti instability strip
and they very likely belong to this class of variables. We notice however 
that no $\delta$ Scuti candidate cooler than F0 was found.

Very little can be said on longer period variables  but it
is worth making some speculation on their possible nature. The observed variability is very likely due to intrinsic changes in 
luminosity of the sources. Indeed a spurious nature due to an observational 
effect, e.g. air transparency variations, would unlikely pass the filter of 
the statistics used: each star is checked against five other selected 
nearby stars in the cluster. In addition if they were due to an observational 
effect they would have been spread out in the HR diagram. They are however located in a narrow color range.

They could be binaries: there is no way to rule out this hypothesis 
through our survey data. We notice however that there is again
no reason why in a bias-free sample of stars the binaries should be 
concentrated on the cold part of the instability strip as they appear.

Could they be intrinsic (pulsating, spotted) variables ? Their position in 
HR diagram and their apparent time-scale suggests they could be $\gamma$ Dor 
star. 

The $\gamma$ Dor stars are variables in time scales between $0-5$ $d^{-1}$ and
with typical amplitudes of $0^m.03-0^m.1$ lying in a very narrow region
on or just beyond the cool border of the instability strip. Even if some 
controversy is still present, they have been recognized as the first evidence 
of non radial $g$-mode pulsation among F-type dwarfs. The position in HR 
diagram of the most thoroughly studied $\gamma$ Dor stars is reported 
(triangles) in fig. 2 for comparison.
  
Not much of an interpretation can be worked out from survey data of the kind 
presented in this paper and the possible $\gamma$ Dor nature of the longer 
period variables should be regarded as guessed.

\section{X-ray data}

The Interaction between pulsation and chromospheric or coronal heating 
mechanisms has been recently indicated as a possible important tool
to collect information both on pulsation mechanisms and heating.

Known single X-ray sources among early F-type stars are generally related
with emission from a corona heated by magnetic fields. Two important cases
are reported in the literature: 47 Cas (G\"udel et al., 1995) and 
71 Tau (Stern et al 1994). The former is reported to be a powerful 
X-ray source showing pseudo-regular variability on a time scale of the order of 
1 d. Its Str\"omgren photometry (Olsen 1983) places it in the $\gamma$ Dor 
region of the HR diagram, and the time scale falls in the typical range of 
the $\gamma$ Dor optical variability. However, recent photometric observations
exclude light curve variability of 47 Cas larger than 4 mmag (Poretti, private
communication). 71 Tau is the strongest X--ray emitter in the Hyades; it 
is a $\delta$ Scuti pulsator with spectral type A9 and large rotational
velocity. It is a binary star, but it is difficult to explain the intense
X--ray emission only with a late--type companion. 

The simultaneous availability of our complete survey and ROSAT data on 
NGC 2516 prompted us to look for relations between variables and X-ray 
sources.

The open cluster NGC 2516 was observed with the Position Sensitive
Proportional counter (PSPC) at the focus of the X--ray telescope on board
ROSAT. This satellite contains an X--ray telescope with a $2^o$
field of view. The PSPC is a gas-filled proportional counter sensitive in
the energy band 0.1--2.4 keV, with a spatial resolution of $\sim 25''$ in
the center of the focal plane. A detailed description of the satellite and
detectors can be found in Tr\"umper (1983) and Pfeffermann et al. (1986). 

The observation reported here was obtained on 1992 April 6 
with a total effective exposure time of 9284 s. Due to Earth occultation,
radiation belt passages and observations of other targets, the data 
were spread over two days. 
The data were analyzed with the XANADU package: first the event files
were taken from the ROSAT on-line archive at the Max Planck Institute
(Munich) and images were extracted. The exposure maps were linearly 
interpolated and rescaled so as to be overlaid on the images. A boresight 
correction of $\sim 10"$ was applied. Exposure corrected count rates or 
$3\,\sigma$ upper limits were finally derived (see tab.~3).

\begin{table*}
\begin{center}
\begin{tabular}{l r r | l r r}
Star & Count rate or $3\,\sigma$ upper limit& S/N&
Star & Count rate or $3\,\sigma$ upper limit& S/N\\
\hline
C4  &$(3.86 \pm 0.81)\times 10^{-3}$ & 4.781  &
C11 &$2.17\times 10^{-3}$            &        \\
C17 &$(3.93 \pm 0.78)\times 10^{-3}$ & 5.031  & 
C21 &$2.47\times 10^{-3}$            &        \\
C22 &$3.31\times 10^{-3}$            &        &
C28 &$1.28\times 10^{-3}$            &        \\
C32 &$1.52\times 10^{-3}$            &        &
C38 &$(2.28 \pm 0.65)\times 10^{-3}$ & 3.503  \\
C40 &$1.23\times 10^{-3}$            &        &
C42$^S$ &$2.47\times 10^{-3}$        &        \\
C44 &$1.08\times 10^{-3}$            &        &
C45 &$3.19\times 10^{-3}$            &        \\
C46 &$1.08\times 10^{-3}$            &        &
C47 &$1.39\times 10^{-3}$            &        \\
C50 &$1.72\times 10^{-3}$            &        &
C51$^S$ &$1.19\times 10^{-3}$        &        \\
C52$^L$ &$1.09\times 10^{-3}$        &        & 
C53 &$1.06\times 10^{-3}$            &        \\
C54$^S$ &$2.44\times 10^{-3}$        &        & 
C55$^S$ &$1.43\times 10^{-3}$        &        \\
C56 &$1.64\times 10^{-3}$            &        &
C59 &$1.80\times 10^{-3}$            &        \\
C60 &$1.41\times 10^{-3}$            &        & 
C62$^L$ &$2.30\times 10^{-3}$        &        \\
C63 &$2.82\times 10^{-3}$            &        &
C64$^L$ &$3.51\times 10^{-3}$        &        \\
C65 &$1.50\times 10^{-3}$            &        & 
C66 &$1.95\times 10^{-3}$            &        \\
C67 &$3.27\times 10^{-3}$            &        & 
C68 &$1.71\times 10^{-3}$            &        \\
C69 &$2.34\times 10^{-3}$            &        &
C77 &$(1.96 \pm 0.61)\times 10^{-3}$ & 3.187  \\
C84$^S$ &$1.26\times 10^{-3}$        &        & 
C92$^L$ &$1.96\times 10^{-3}$        &        \\
C93$^L$ &$1.74\times 10^{-3}$        &        &
C96 &$2.43\times 10^{-3}$            &        \\
C97 &$1.66\times 10^{-3}$            &        &
C98 &$(3.08 \pm 0.79)\times 10^{-3}$ & 3.893  \\
C99 &$3.28\times 10^{-3}$            &        &
C101&$1.74\times 10^{-3}$            &        \\
C106$^L$&$1.46\times 10^{-3}$        &        &
C107&$2.58\times 10^{-3}$            &        \\ 
C108&$2.29\times 10^{-3}$            &        &
    &                                &        \\
\hline
\end{tabular}
\end{center}
\caption{Detection of X--ray source in NGC 2516: the label $^S$ or $^L$ 
indicates variable type according to the classification reported in the text.
count rate or upper limits are expressed in count s$^{-1}$.}
\end{table*}

The analysis showed that none of the suspected variables in the cluster
has an X--ray emission (0.1--2.4 keV) above $\sim 5\times 10^{29}$ erg
s$^{-1}$ (assuming a count rate to flux conversion factor of $1 {\rm CR} =
10^{-11}$ erg s$^{-1}$ cm$^{-2}$ and a cluster distance of 440 pc).
This upper limit has to be compared with $\sim 10^{31}$ erg s$^{-1}$ for 
47 Cas (G\"udel et al. 1995) and $\sim 3\times 10^{30}$ erg s$^{-1}$ for 
71 Tau (Stern et al 1992). It is interesting to note that, despite the 
non-detection of variable stars, some stars inside the variability strip 
have been revealed with an X--ray luminosity of a few $10^{29}$ 
erg s$^{-1}$. 

\section{Conclusions}

In this paper we have reported on a survey for variability of a selected
sample of A- and early F-type stars in the southern open cluster NGC 2516.
The sample is complete in the sense that all dwarfs and subgiants with
$0.00<(b-y)_o<0.25$ belonging to the cluster have been observed. 

The new results confirm the normal incidence of shorter period variables 
(likely $\delta$ Scuti) and the anomalous incidence of longer period 
variables previously found. The results have been also compared to the 
survey in the X band obtained by ROSAT but no relation between variables 
and X-ray sources could be found.

The statistic used is designed to be less sensitive to spurious effects such as
air transparency variations. Observational effects can therefore be excluded
and the variability ascribed to variations of the luminosity of the sources.

While the hypothesis of binarity for explaining longer period variables can not be ruled out through our
data, the position occupied by them suggests 
their possible belonging to the newly discovered class of $\gamma$ Dor 
variables.
 
If confirmed, the $\gamma$ Dor nature of the longer period variables in NGC 
2516 would be an important result because it is unique among clusters, and the 8 objects disentangled in our survey certainly
deserve further investigation.

\end{document}